\documentclass[prl,onecolumn,showpacs,showkeys,superscriptaddress]{revtex4}

\usepackage{amssymb,amsmath,graphicx}
\begin{document}

\title{Enhanced sequential carrier capture into individual quantum dots and quantum posts controlled by surface acoustic waves}
\author{Stefan V\"olk}
\altaffiliation{These authors contributed equally to this work.}
\affiliation{Lehrstuhl f\"ur Experimentalphysik 1, Universit\"at Augsburg, Universit\"atsstr. 1, 86159 Augsburg, Germany}
\author{Florian J. R. Sch\"ulein}
\altaffiliation{These authors contributed equally to this work.}
\affiliation{Lehrstuhl f\"ur Experimentalphysik 1, Universit\"at Augsburg, Universit\"atsstr. 1, 86159 Augsburg, Germany}
\author{Florian Knall}
\affiliation{Lehrstuhl f\"ur Experimentalphysik 1, Universit\"at Augsburg, Universit\"atsstr. 1, 86159 Augsburg, Germany}
\author{Dirk Reuter}
\affiliation{Lehrstuhl f\"ur Angewandte Festk\"orperphysik, Ruhr-Universit\"at Bochum, Universit\"atsstr. 150, 44780 Bochum, Germany}
\author{Andreas D. Wieck}
\affiliation{Lehrstuhl f\"ur Angewandte Festk\"orperphysik, Ruhr-Universit\"at Bochum, Universit\"atsstr. 150, 44780 Bochum, Germany}
\author{Tuan A. Truong}
\affiliation{Materials Department, University of California, Santa Barbara CA 93106, United States}
\author{Hyochul Kim}
\affiliation{Physics Department, University of California, Santa Barbara CA 93106, United States}
\author{Pierre M. Petroff}
\affiliation{Materials Department, University of California, Santa Barbara CA 93106, United States}
\author{Achim Wixforth}
\affiliation{Lehrstuhl f\"ur Experimentalphysik 1, Universit\"at Augsburg, Universit\"atsstr. 1, 86159 Augsburg, Germany}
\author{Hubert J. Krenner}
\email{hubert.krenner@physik.uni-augsburg.de}
\affiliation{Lehrstuhl f\"ur Experimentalphysik 1, Universit\"at Augsburg, Universit\"atsstr. 1, 86159 Augsburg, Germany}

\begin{abstract}

Individual self-assembled Quantum Dots and Quantum Posts are studied under the influence of a surface acoustic wave. In optical experiments we observe an acoustically induced switching of the occupancy of the nanostructures along with an overall increase of the emission intensity. For Quantum Posts, switching occurs continuously from predominantely charged excitons (dissimilar number of electrons and holes) to neutral excitons (same number of electrons and holes) and is independent on whether the surface acoustic wave amplitude is increased or decreased. For quantum dots, switching is non-monotonic and shows a pronounced hysteresis on the amplitude sweep direction. Moreover, emission of positively charged and neutral excitons is observed at high surface acoustic wave amplitudes. These findings are explained by carrier trapping and localization in the thin and disordered two-dimensional wetting layer on top of which Quantum Dots nucleate. This limitation can be overcome for Quantum Posts where acoustically induced charge transport is highly efficient in a wide lateral Matrix-Quantum Well.
\end{abstract}
\maketitle

\begin{figure}[htbp]
\begin{center}
\includegraphics[width=0.95\columnwidth]{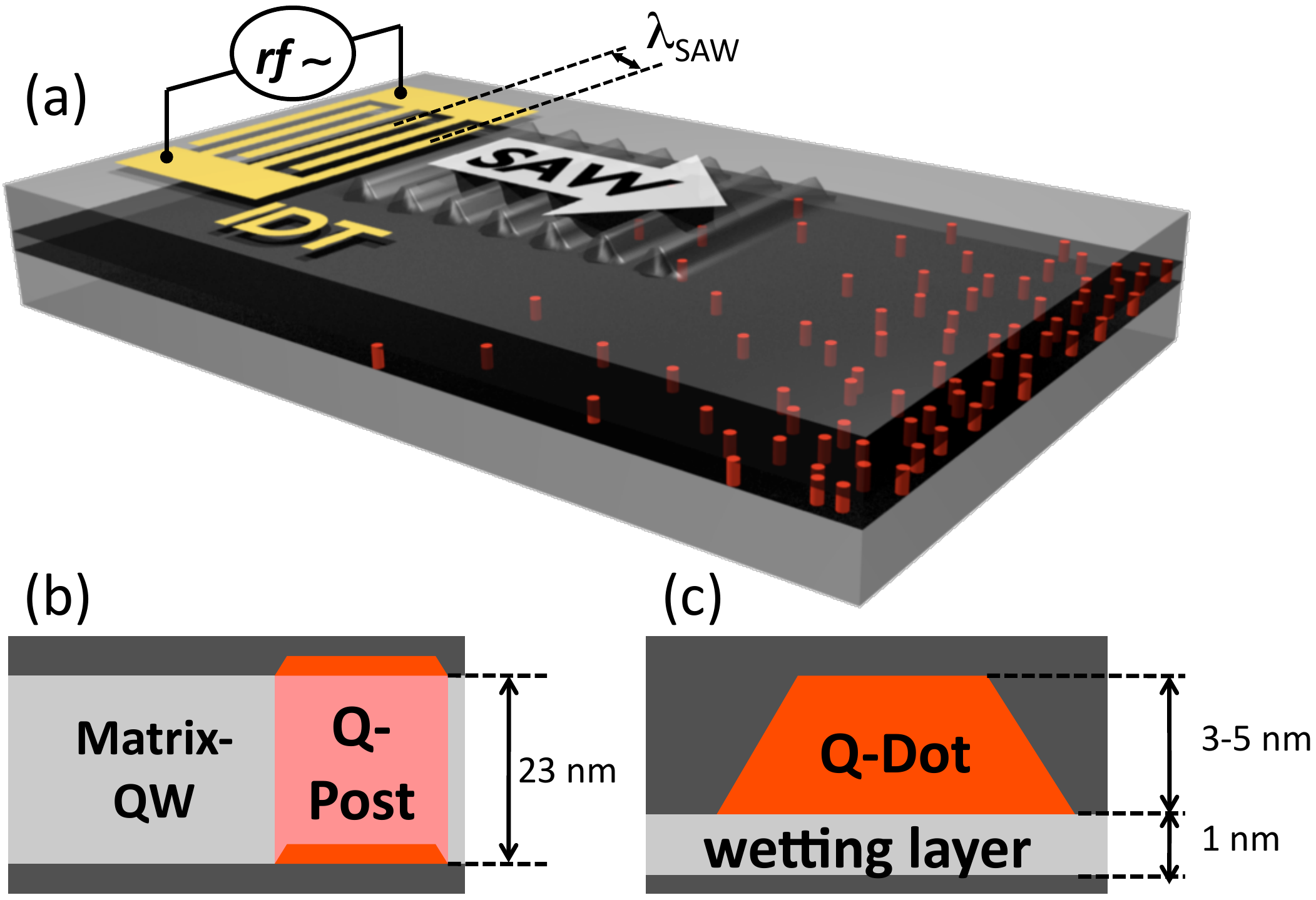}
\caption{(a) Schematic of a device consisting of a density gradient of Q-Posts and an interdigital transducer (IDT) to excite SAWs. (b) and (c) schematic of a Q-Post and a Q-Dot.}

\label{Fig:1}
\end{center}
\end{figure}

For almost 20 years, the zero-dimensional confined levels of quantum dot nanostructures (Q-Dots) have been studied intensely for both novel optoelectronic device applications but also for more advanced concepts in the field of quantum information processing and communication\cite{Michler:09:book,Gywat:09:book}. A key requirement for these applications is the control of the number of carriers occupying the Q-Dot as well as the emission and interaction energies of  the respective charge and spin excitations. In voltage tunable field-effect structures, Q-Dots are coupled to a reservoir from which a well-defined number of carriers can be injected\cite{Drexler:94,Reuter:05}. This approach is employed for electrical capacitance-voltage spectroscopy of electron and hole levels of the Q-Dots and can be adopted for optical methods to study charged excitons in Q-Dots\cite{Warburton:00,Finley:04} and tunable artificial Q-Dot-molecules\cite{Krenner:05b,Stinaff:06,Robledo:08}.\\
Since the early 1980'ies surface acoustic waves (SAWs) have been employed to probe fundamental physical effects in low-dimensional semiconductor structures such as the Quantum Hall Effects\cite{Wixforth:86,Kukushkin:09} or to transport individual charges and spins\cite{Rotter:99a,Sogawa:01}. In optically active structures the acousto-electric fields induced in piezoelectric semiconductors lead to a dissociation and spatial separation of electron-hole pairs (excitons) in a quantum well (QW). This gives rise to a charge conveyance effect \cite{Rocke:97} which can be further employed to realize e.g. an acoustically triggered single photon source \cite{Wiele:98,Boedefeld:06,Couto:09}.\\
In piezoelectric materials such as $\mathrm{LiNbO3}$, $\mathrm{ZnO}$ or compound III-V semiconductors SAWs can be excited directly by pure electrical means using so-called interdigital transducers (IDTs). When a radio frequency (RF) voltage is applied to the two opposite electrodes, a SAW of wavelength $\lambda_{\rm SAW}$ and frequency $f_{\rm SAW}$ is excited if the periodicity of the IDT and the applied RF frequency match the SAW dispersion relation $f_{\rm SAW} =c_{\rm sound}/ \lambda_{\rm SAW}$ as depicted schematically in \ref{Fig:1} (a). Since SAWs are acoustic excitations, $c_{\rm sound}$ is the velocity of sound of the material which for $\mathrm{GaAs}$ is $\sim 3000\mathrm{~m/s}$. Using IDTs fabricated by standard optical and electron beam lithography SAWs cover frequencies over a wide band from hundreds of kHz up to several GHz. SAWs propagate almost without dissipation over macroscopic distances exceeding several millimeters and, therefore, a large number of nanostructures located within its propagation path are addressed by the electrically generated SAW. Moreover, individual nanostructures can be independently identified and studied under the perturbation of the SAW using all-optical techniques. This \emph{massively parallel} high-frequency addressing and manipulation by SAW is in strong contrast to field-effect devices where similar electrical switching with speeds $>1~\rm GHz$ require sophisticated nanofabricated contacts to \emph{individual}, single Q-Dots due to the capacitance of these device. \\
Here, we present a combined photoluminescence study on self-assembled Q-Dots and Quantum Posts (Q-Posts)\cite{He:07,Li:08b} subject to the propagating piezoelectric potential induced by a high frequency SAW. We demonstrate a pronounced change of the occupancy state of the Q-Dots and Q-Post as the SAW power is increased manifesting itself in a switching between neutral and charged excitons in the time-integrated emission spectra. The observed behavior is explained by a SAW-driven change of the nature of the carrier capture into the confined levels of Q-Dots and Q-Posts. It arises from a regime of simultaneous injection of both electrons and holes at low SAW amplitudes to regime at higher amplitudes in which the two carrier species are injected sequentially. The distinct differences of this switching between Q-Dots and Q-Posts is attributed to a SAW-driven dissociation of electron-hole pairs within the two-dimensional QWs coupled to the respective nanostructure.\\

The two types of samples studied in this work have been grown by molecular beam epitaxy (MBE) and contain a single layer of either self-assembled Q-Dots or Q-Posts. The two types of nanostructures studied differ inherently in their morphology and their adjacent two-dimensional QWs as sketched in \ref{Fig:1} (b) and (c). Conventional Q-Dots are flat, 3 - 5 nm high, In-rich islands with a diameter in the range of 20 nm which nucleate on a thin $\sim1$ nm InGaAs wetting layer (WL) resembling a narrow QW with a larger effective bandgap than the Q-Dots themselves. Self-assembled Q-Posts on the other hand are columnar structures with similar diameters as the Q-Dots. However, their height can be adjusted with nanometer precision from $\sim3$ nm to > 60 nm while preserving the high optical quality of conventional Q-Dots \cite{Krenner:08a,Krenner:09}. Q-Posts themselves are embedded laterally in a $d_{\rm ~Matrix}$ wide $\mathrm{InGaAs}$ Matrix-QW with a reduced $10\%$ In-content. This ensures full, quasi-zero-dimensional carrier confinement in the Q-Post. The Q-Posts studied in these experiments are fabricated by eight repetitions of the employed deposition sequence \cite{He:07} and have a nominal height of 23 nm, corresponding to an aspect ratio (height/diameter) of $\sim1$. Accordingly, the Matrix-QW has the same thickness of $d_{\rm~Matrix} = 23$ nm. For both samples, a gradient of the Q-Dots or Q-Posts surface density was realized by intentionally stopping the substrate rotation during MBE growth of these layers\cite{Leonard:94,Krenner:05a}. The position at which the transition to low surface density occurs was determined by conventional scanning micro-photoluminescence ($\mu$-PL) at low temperatures on each wafer. IDTs were aligned with respect to this surface density gradient such that the region with less than one nanostructure per $\mu \rm m^{2}$ was in the propagation direction of the SAW. Details on this procedure and the sample layout may be found in the Supporting Information. A sketch of the full sample structure on which individual Q-Dots and Q-Posts can be studied by $\mu$-PL under the influence of SAWs is depicted in \ref{Fig:1} (a). For the experiments presented here we used IDTs with wavelengths of $\lambda_{\rm ~Q-Dot}=11.6\mathrm{~\mu m}$ and $\lambda_{\rm~Q-Post}=15\mathrm{~\mu m}$ for the Q-Dot- and Q-Post-sample, respectively, which correspond to excitation frequencies of $f_{\rm ~Q-Dot} = 251.5$ MHz and $f_{\rm ~Q-Post} = 193$ MHz at $T = 4$ K for the two designs. \\
All $\mu$-PL experiments were performed at low temperatures ($T=4~\mathrm{K}$) with the sample mounted on the coldfinger of an optical He flow-cryostat. Carriers were photogenerated using an externally triggered diode laser providing < 100 ps pulses at $\lambda = 661$ nm focused to a $\sim 2~ \mu\rm m$ spot using a $50\times$ microscope objective. Weak optical excitation $(\sim 100$ nW) was employed in order to excite only single $s$-shell excitons and avoid multi-exciton generation. The emission from the sample was collected by the same objective lense, dispersed using a 0.5 m grating monochromator and detected by a \emph{l}$\mathrm{N2}$ cooled Si-CCD detector with an overall resolution $< 150~\mu\rm eV$. The influence of the SAW on the Q-Dot or Q-Post emission was probed as a function of the RF-power applied to the IDT, whereof the square root is directly proportional to the amplitude of the excited SAW. In order to exclude heating of the sample at high power levels, we used \emph{pulsed} SAW excitation with a repetition frequency of 100 kHz and an "on"-"off" duty cycle of $\sim1$:9. The excitation laser was synchronized to the $\sim1~\mathrm{\mu s}$ long SAW pulses and the $\mu$-PL signal was integrated over several tens of seconds. Further details of this pulse scheme are provided in the Supporting Information.\\

\begin{figure}[htbp]
\begin{center}
\includegraphics[width=0.95\columnwidth]{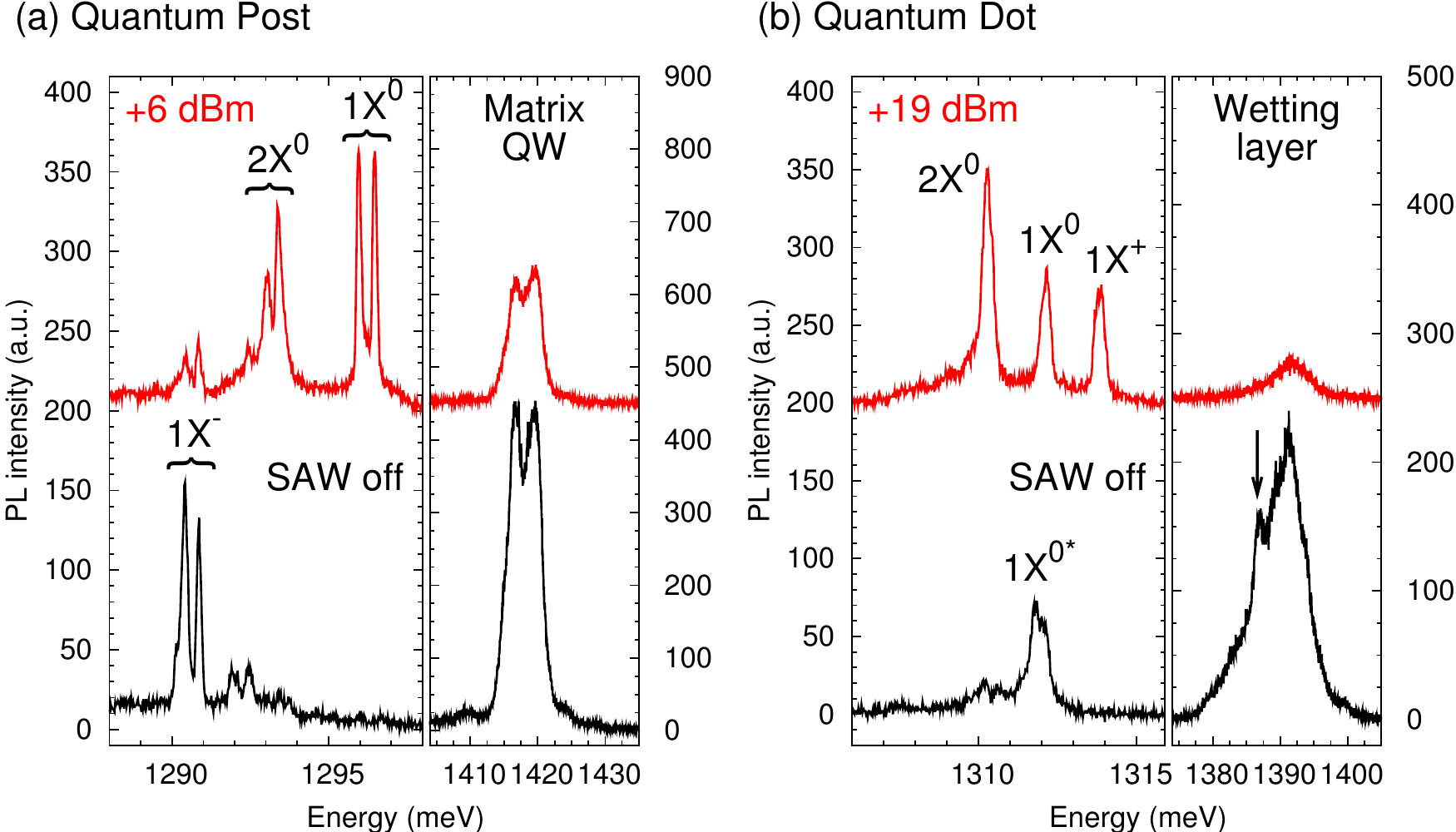}
\caption{Emission spectra of a single Q-Post and Matrix-QW (a) and a single Q-Dot and WL(b) without and under the influence of a SAW. With the SAW applied, a clear switching of emission lines and an overall increase of intensity is observed.}

\label{Fig:2}
\end{center}
\end{figure}

We start the description of our experimental findings by comparing the emission spectra of the Q-Dot and Q-Post samples without and with a SAW applied. Two representative examples of each system are shown in \ref{Fig:2}. When no SAW is excited (\ref{Fig:2} (a) black, lower spectra) we observe PL from the Q-Post and a strong signal from the Matrix-QW. The emission spectra undergo a pronounced change when the SAW is turned on (red, upper spectra): The Matrix-QW signal shows the expected quenching behavior at $P_{\rm ~SAW}$ = +6 dBm, indicative for charge conveyance\cite{Rocke:97}. This is further confirmed by an increase of the conductivity in the sample as independently measured by the change of transmission of SAWs \cite{Rocke:98}. The Q-Post shows a clear exchange of intensity between an emission doublet at $E_{1X^-} = 1290.5$ meV emitting without SAW and a new doublet at $E_{1X^0} = 1296.2$ meV with a weak overall increase of emission intensity. These two doublets are attributed to recombinations of the negatively charged $(1X^- = 2e + 1h)$ and the charge neutral $(1X^0 = 1e + 1h)$ exciton in the Q-Post. We want to note that - in contrast to conventional Q-Dots - single Q-Posts frequently show emission doublets arising from the two different \emph{localized} hole levels at the two ends of the Q-Post \cite{Krenner:08b}. The relative detuning of these two hole levels is sensitive to the local composition and strain in the Q-Post. For a specific Q-Post this detuning influences the population probability i.e. the intensity ratio of the two lines of the doublet. In the specific sample studied we predominantly observe emission of $1X^-$ without SAWs applied. We want to note that an analogous switching behavior is also observed for Q-Posts for which emission of both $1X^0$ and $1X^{-}$ is detected in time-integrated spectra without a SAW present and an example is discussed in the Supplementary Information. In contrast to a Q-Post, a conventional Q-Dot [c.f. \ref{Fig:2} (b)] exhibits a single emission line without SAW (black, lower spectrum). This relatively broad emission line at $E_{1X^{0*}} = 1311.7$ meV splits with $P_{\rm ~SAW}$ = +19 dBm into three individual lines at $E_{2X^0} = 1310.3$ meV, $E_{1X^0} = 1312.1$ meV and $E_{1X^{+}} = 1313.8$ meV with strong increase of overall intensity by $\sim 1$ order of magnitude, while the WL PL is quenched almost completely. From the observed energetic splittings we attribute these emission lines to arise from radiative transitions from different $s$-shell excitons localized in the Q-Dot: the charge neutral exciton $(1X^0)$, biexciton $(2X^0 = 2e + 2h)$ and positively charged exciton $(1X^{+}=1e + 2h)$. The observed splittings and attributions for both Q-Posts and Q-Dots are further supported by power-dependent PL spectroscopy and consistent with previous experimental and theoretical studies of various Q-Dot systems and in particular for self-assembled $\mathrm{In(Ga)As}$ Q-Dots emitting in this energy range \cite{Warburton:00,Finley:04,Brunner:94a,Bayer:00,Regelman:01,Scheibner:07b,Schuelein:09,Schliwa:09}. Clearly, with a SAW applied for both systems the spectra are dominated by the neutral excitons $1X^{0}$ and $2X^0$. However, in contrast to the Q-Post, where charged species are almost completely suppressed, strong emission of the \emph{positively} charged exciton $1X^{+}$ is observed in the time-integrated spectrum of a Q-Dot. Moreover, for the Q-Dot, the emission of the neutral single exciton is initially broadened significantly ($1X^{0*}$) and narrows when the SAW is turned on ($1X^{0}$).\\

\begin{figure}[htbp]
\begin{center}
\includegraphics[width=0.75\columnwidth]{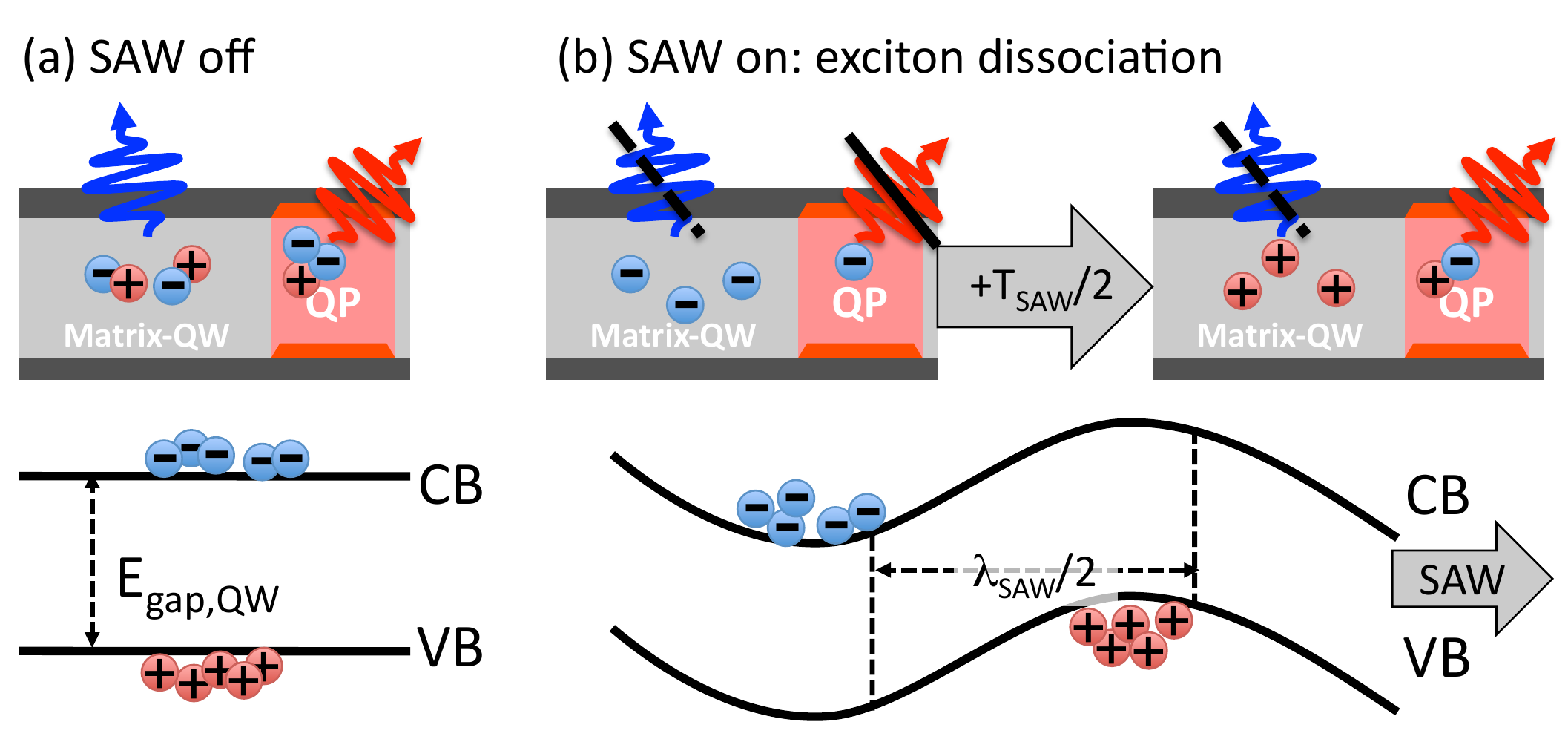}
\caption{Random vs. sequential carrier injection schematically shown for a Q-Post and its Matrix-QW: (a) Without SAWs, electrons and holes are present at the same time at the position of the nanostructure and emission of the Q-Post and the Matrix-QW is observed. (b) With SAWs present, electrons and holes are spatially separated and sequentially injected giving rise to exciton recombination. The Matrix-QW emission is largely suppressed over the entire SAW cycle. {\it Lower part:} Schematic of the SAW-induced band edge modulation giving rise to a separation of electrons and holes into local minima and maxima in the conduction (CB) and valence band (VB). These carrier pockets propagate with the speed of sound along with the SAW leading to charge conveyance.}
\label{Fig:3}
\end{center}
\end{figure}

The effect of the SAW on the PL emission can be qualitatively understood by the Type-II band edge modulation caused by the large piezoelectric fields accompanying the SAW. These acousto-electric fields lead to a dissociation of excitons into individual electrons and holes which are localized in the maxima and minima of the SAW, respectively\cite{Rocke:97}. The two different cases are depicted schematically in \ref{Fig:3} for a Q-Post. With no SAW applied [c.f. \ref{Fig:3} (a)] both carrier species are photogenerated and remain at the position of the nanostructure. Thus, bound electron-hole pairs as well as individual carriers can be captured into the Q-Post which in the case of our sample predominantly favors the formation of $1X^{-}$. As the SAW modulates the band edges (lower part of \ref{Fig:3}), the aforementioned charge separation by $\lambda_{\rm~SAW}/2$ occurs and depending on the local phase of the SAW at the Q-Post position \emph{either} electrons or holes are present. This leads to an unique \emph {inherently sequential} injection of the two carrier species: In the example sketched in \ref{Fig:3} (b), first electrons are present in the Matrix-QW and become captured in the Q-Post. After one half period $(T_{~\rm SAW}/2)$ of the SAW later, the opposite carrier species, in this case holes, arrive. Since at this stage the Q-Post already contains electron(s) with a net negative charge, the attractive Coulomb interaction favors capture of the same number of holes giving rise to the observed preferential generation of charge neutral excitons. These excitons then recombine radiatively by emission of single photons, while the Matrix-QW signal is largely suppressed over the entire SAW cycle. This model, however, cannot describe the observation of both neutral and charged excitons for Q-Dots in our time-integrated measurements and the different power levels at which this effect occurs for the two types of nanostructures. \\

\begin{figure}[htbp]
\begin{center}
\includegraphics[width=0.95\columnwidth]{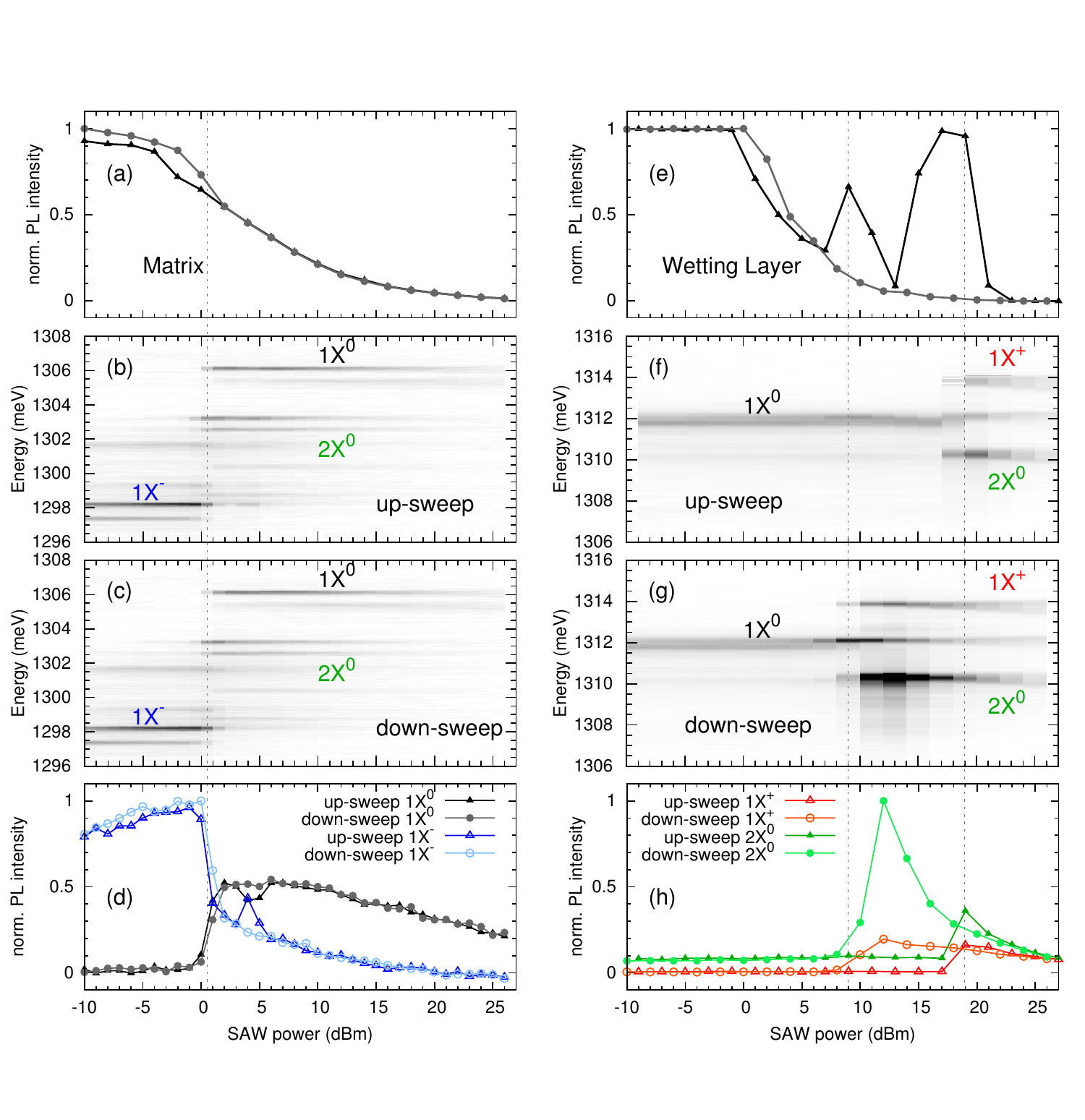}
\caption{Comparison of SAW power sweeps for a single Q-Post and Q-Dot: Emission intensity of the Matrix-QW (a) and WL (e). Grayscale plots of the Q-Post (b), (c) and Q-Dot (f), (g) emission for SAW up- and down-sweeps. Emission intensities of different emission lines of the Q-Post (d) and Q-Dot (h) [triangles: up-sweep, circles: down-sweep].}
\label{Fig:4}
\end{center}
\end{figure}

In order to investigate the underlying mechanism and the differences between the Q-Post and Q-Dot sample in detail, we studied this effect over a wide range of $P_{\rm ~SAW}$  from -10 dBm to +26 dBm. In \ref{Fig:4}, we present a complete overview of $\mu$-PL data from a Q-Post and Q-Dot and the respective Matrix-QW and WL signals for both increasing (up-sweep) and decreasing (down-sweep) SAW power. Integrated PL-intensities are plotted as triangles and circles for the up- and down-sweep, respectively.\\
We start by examining the integrated and normalized PL intensity of the Matrix-QW of the Q-Post sample which is plotted in \ref{Fig:4} (a). The signal exhibits the characteristic quenching behavior of a QW as $P_{\rm ~SAW}$ is increased\cite{Rocke:97}. This quenching of the Matrix-QW is strictly monotonic for both the up- and down-sweep and sets in at $P_{\rm ~SAW} = -8 \rm ~dBm$. This continuous behavior of the Matrix-QW reflects itself also in the emission of single Q-Posts. A representative example of an up- and down-sweep is plotted over the same range of $P_{\rm ~SAW}$ in grey-scale representation in panels (b) and (c) of \ref{Fig:4}, respectively. In the detailed analysis of this data [\ref{Fig:4} (d)] two regimes can be distinguished: (i) For $  -10{\rm~dBm}<P_{\rm ~SAW} <0 ~{\rm dBm}$ the overall emission intensity increases weakly and (ii) for further increasing the $P_{~\rm SAW}> +0.5\rm dBm$ we observe a clear switching between emission from $1X^{-}$ at low SAW powers to $1X^{0}$ and $2X^{0}$ at high levels of $P_{\rm ~SAW}$. In the low SAW power regime the induced band edge modulation is not sufficient to induce charge conveyance and as $P_{\rm ~SAW}$ increases the induced piezoelectric fields dissociate a larger fraction of the photogenerated bound electron-hole pairs which are then captured as individual electrons and holes into the confined Q-Post levels. This capture of individual carriers is more efficient than that of bound pairs and this effect has been observed in temperature dependent experiments using thermal dissociation as the driving mechanism \cite{He:07,Krenner:09}. Moskalenko {\it et al.}\cite{Moskalenko:07} observed a similar effect using static lateral electric fields to dissociate photogenerated excitons. For higher SAW power levels charge conveyance sets in giving rise to a \emph{sequential} injection of electrons and holes as sketched in \ref{Fig:3}. Since switching occurs within a very narrow $(\pm 1\rm ~ dBm)$ range around $P_{\rm ~SAW} = +0.5$ dBm in \emph{both} the up- and the down-sweep [vertical dashed line in \ref{Fig:4} (a-d)] the transition from the carrier dissociation to charge conveyance regime is continuous and the SAW-induced piezoelectric fields are only screened by the photogenerated carriers.\\ 

We performed the same detailed SAW power sweeps for the WL and Q-Dot signals which are presented in \ref{Fig:4} (e-h). We find pronounced differences compared to the Q-Post and Matrix-QW emissions: The WL of the Q-Dot sample [\ref{Fig:4} (e)] shows that the onset of the SAW-induced quenching is increased to $P_{\rm ~SAW} = -1$ dBm (corresponding to a factor of 5) compared to the Matrix-QW. Most strikingly, the quenching behavior itself is \emph{non}-monotonic for the up- and strictly monotonic for the down-sweep. This oscillatory behavior of the WL emission intensity during the up-sweep persists over two orders of magnitude of RF-power, until from $P_{\rm ~SAW} = +19$ dBm onwards the signal is completely suppressed. When starting from these high power levels, the quenching persists down to $P_{\rm ~SAW} = +12$ dBm and, for further decreasing $P_{\rm ~SAW}$, the original signal level is restored strictly monotonically. The Q-Dot emission [c.f . \ref{Fig:4} (f) and (g)] exhibits also an hysteretic switching in the up- and down-sweep: The characteristic splitting into three emission lines of $1X^{+}$, $1X^0$ and $2X^0$ occurs in the up-sweep at exactly the same power level at which the oscillatory behavior of the WL emission breaks down. When $P_{\rm ~SAW}$ is reduced, this characteristic pattern persists down to lower levels and switches back to a single emission line at exactly the level at which the QW signal recovers as marked by the vertical dashed lines at $P_{\rm ~ SAW} = +9$ dBm and $+19$ dBm in \ref{Fig:4} (e-h). We want to note that the two switching points for the up- and down-sweeps are split by $\sim 10$ dB corresponding to a variation of $P_{\rm ~SAW}$ by one order of magnitude. Furthermore, all lines show a pronounced broadening at high values of $P_{\rm ~SAW}>+20$ dBm due to piezoelastic and Stark shifts\cite{Gell:08,Finley:04}. Whilst the latter effect is also present for Q-Posts, the Q-Dot switching shows three major differences: (i) emission of the \emph{positively} charged exciton in addition to neutral excitons at high SAW powers, (ii) stronger increase of intensity at the switching point, and (iii) switching occurs at different levels of $P_{\rm ~SAW}$ in the up and the down-sweep [$P_{\rm ~SAW, up } >P_{\rm ~SAW,down}$]. These facts are further illustrated in \ref{Fig:4} (d) and (h), where, as examples, the intensities of the Q-Post $1X^0$ (closed symbols) and $1X^{-}$ (open symbols) and Q-Dot $1X^{+}$ (open symbols) and $2X^0$ (closed symbols) emissions are summarized. In this analysis, the strong (weak) increase of the overall intensity and (non-)hysteretic switching is clearly visible for the Q-Dot (Q-Post).\\

The drastically different  behaviors can be understood by considering the morphology of the two samples in study. Here, the major difference arises from the two-dimensional systems which these nanostructures are coupled to. Carriers are photogenerated in our experiments in the entire GaAs host matrix and relax via WL and Matrix-QW states into the energetically favorable Q-Dot and Q-Post levels, respectively. The Matrix-QW laterally surrounding the Q-Posts is a wide potential well (in our case $d_{\rm ~Matrix}=23$ nm) while the WL on top of which Q-Dots nucleate is thin and disordered. This leads to a pronounced difference in the dissociation of excitons by the SAW-induced electric fields and their charge conveyance. The exciton binding energy is lower for a wider QW and, thus, lower SAW amplitudes are required to overcome this energy and to establish charge conveyance in the wide Matrix-QW compared to the thin WL. In addition, for thin QWs, width variations lead to local potential fluctuations such as "natural" Q-Dots  leading to localization of both excitons and individual charges\cite{Babinski:08,Voelk:09,Fuhrmann:10a}. Such localized charges are observed as sharp spectral features in the WL signals as marked by an arrow in \ref{Fig:2} (b). Such fluctuations are expected to be enhanced for the thin WL resulting in the observed pronounced intensity oscillations in this power range due to partial screening of the piezoelectric potentials by trapped carriers. After the onset of exciton dissociation in the WL in the range of -1 dBm $<P_{\rm ~SAW}< +12$ dBm, electrons become partially mobile while holes remain mostly localized. This has been observed previously by Alsina {\it et al.} in time-resolved studies\cite{Alsina:01} for SAW-driven charge conveyance in QWs. As $P_{\rm ~SAW}$ is increased further to $> +19$ dBm, the SAW amplitude is sufficient to fully remove holes from localization sites and, therefore, the WL emission is quenched and switching occurs in the Q-Dot signal. From this power level onwards efficient charge conveyance \emph{abruptly} sets in, further supported by the complete suppression of the WL emission. More evidence for this interpretation is given in the Supporting Information where we examine the WL PL in more detail. We observe emission from "natural" quantum dots \cite{Babinski:08} localized at interface fluctuations \emph{only} during the down-sweep of the SAW. This indicates that the corresponding localization sites are randomly occupied initially and have to be cleared by the high power SAW before exciton localization can occur. Remarkably, the characteristic pattern of $1X^{+}$, $1X^0$ and $2X^ 0$ persists down to significantly lower SAW powers in the down-sweep compared to the up-sweep. Moreover, a narrowing of emission lines is observed \emph{after} a high SAW power level was applied and then subsequently reduced, whereas with no SAWs applied, the emission is broadened. This broadening could arise e.g. from a described fluctuating charge environment surrounding the Q-Dot which is suppressed or emptied by the high power SAW \cite{Berthelot:06}. Narrow emission lines persist during the down-sweep to lower levels of $P_{\rm ~SAW}$ than required for switching in the up-sweep. This finding leads to the assumption that the build up of fluctuating background charge takes place over timescales exceeding the duration of the experiment i.e. several tens of seconds. For the weak optical pump powers applied in our experiments, the number of photogenerated carriers is not sufficient to quickly drive this charge environment into an equilibrium. This quasi-static background is present initially and has to be considered in the interpretation of the experimental findings: When $P_{\rm ~SAW}$ increases during the up-sweep, the fluctuating charge environment electrically screens the piezoelectric potential induced by the SAW and charge conveyance is inhibited \cite{Rocke:98}. At $P_{\rm ~SAW}=+19~{\rm dBm}$ the potential modulation induced by the SAW exceeds a critical level required to remove the charge environment and the induced screening effect breaks down. This results in the observed abrupt onset of efficient charge conveyance for electrons, whereas for holes the effect is weak. As the SAW power is successively decreased, the efficiency of electron charge conveyance is reduced. Since the charge background is restored slowly over a seconds timescales this reduction is \emph{continuous} during the down-sweep. Charge conveyance smoothly breaks down close to the switching point of the down-sweep at $P_{\rm ~SAW} \sim +10{\rm~dBm}$ where the electron-hole pairs are ionized by band edge modulation but electrons are no longer efficiently spatially separated along the SAW propagation direction. This situation is analogous to Q-Posts in both the up- and down-sweep for which pronounced localization effects do not occur. Moreover, in this power range close to the switching point during the down-sweep both individual electrons and holes are present at the position of the Q-Dot which gives rise to a strong increase of the overall capture probability and, therefore, emission intensity as discussed previously for the case of Q-Posts.\\

In the final part of the discussion of these experimental findings we focus on the observation of $1X^{+}$ emission for Q-Dots in the charge conveyance regime. This effect arises from a net increase of hole over the electron capture probability which can be qualitatively understood by considering the different mobilities of the two carrier species. First of all, we expect a strong reduction of the carrier mobilities in the WL compared to the Matrix-QW due to interface roughness effects similar to observations of two-dimensional electron systems in QWs \cite{Sakaki:87}. In addition, these localization effects and the overall reduction of the mobility are enhanced for holes compared to electrons due to the higher effective hole mass. For electrons (holes), the band edge modulation in the conduction (valence) band induced by the SAW is sufficiently large (too weak) to fully overcome localization effects. Therefore, acousto-electric transport of electrons by the SAW in its propagation direction is enhanced compared to holes \cite{Alsina:01}. Since these different efficiencies of the acoustic transport arise from a reduced mobility of the two carrier species, it is expected to be more pronounced for a thinner QW. Thus, a significantly lower fraction of photogenerated holes is transferred to the SAW-induced valence band maxima for the WL compared to the Matrix-QW while the majority of holes remains stationary at the position of the exciting laser spot. Additionally, electrons (holes) diffuse and spread in the plane of the WL along the conduction band minima (valence band maxima) \emph{perpendicular to the SAW propagation direction} \cite{Krauss:02}. Since the mobility of electrons can exceed the one of holes by more than one order of magnitude, the net hole density and capture probability are increased further with respect to those of electrons at the position of the Q-Dot. Such strong imbalance gives rise to the formation of positively charged excitons. Even the capture of a hole during the radiative cascade from $2X^0$ via $1X^0$ to the crystal ground state cannot be excluded. In contrast to the SAW-conveyed electrons these quasi-immobile holes may relax into the Q-Dot levels during a radiative cascade. The weaker increase of the $1X^+$ intensity compared to $2X^0$ during the down-sweep close to the switching point originates from the \emph{smooth and continuous} break-down of electron charge conveyance and the directional lateral spreading. In this power range, excitons are dissociated but no longer transported by the SAW and both species are present at the Q-Dot position over a full cycle of the SAW. Moreover, electron diffusion is less efficient since it is no longer directional along the SAW-induced conduction band minima\cite{Voegele:09}. Thus, both carrier species are present at the position of the Q-Dot with a net increase of the electron density. Since, in addition, carriers are captured individually in this range of SAW powers, the overall emission intensity increases by more than a factor of 7 with a weaker increase of the $1X^+$ intensity. These effects are expected not to be dominant for a Q-Post with its wide Matrix-QW: the reduced carrier localization leads to a more efficient charge conveyance of both carrier species. In addition, lateral spreading of electrons and holes within their respective, spatially separated conduction and valence band extrema reduces the concentration imbalance. Therefore, simply by Coulomb attraction, a Q-Post containing a fixed number of carriers of one species captures the same number of the oppositely charged carriers upon their arrival at its position. Moreover, we can exclude SAW-assisted electron escape from a Q-Dot into the WL as the origin of the pronounced $1X^{+}$ emission. Since Q-Posts have a confinement potential for electrons and holes similar to Q-Dots, \cite{Krenner:08b} this mechanism would be present in both systems in this SAW power range which is not observed experimentally.\\

\begin{figure}[htbp]
\begin{center}
\includegraphics[width=0.75\columnwidth]{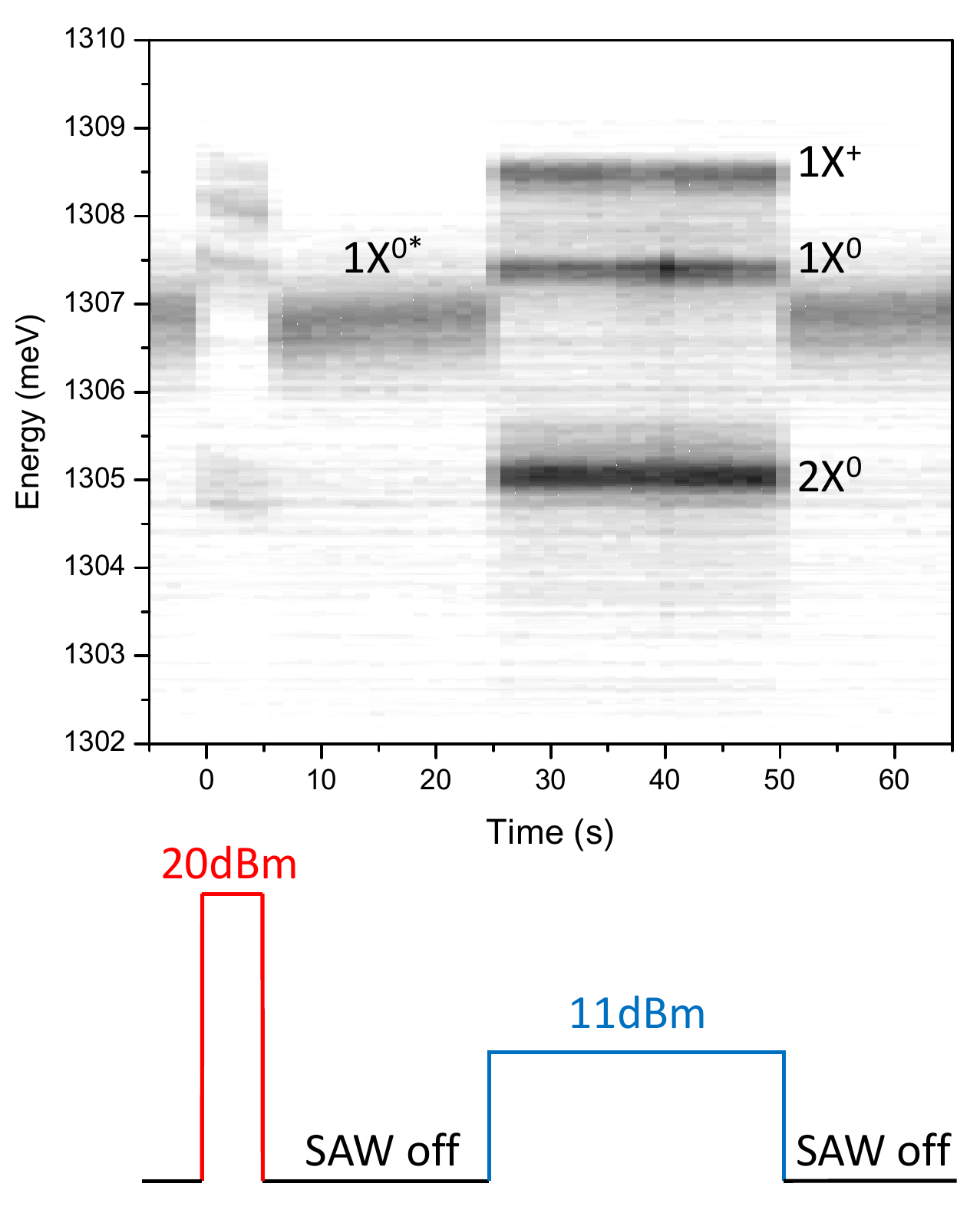}
\caption{Low SAW power switching of a Q-Dot: Emission of the Q-Dot as a function of time. The timing sequence of the SAW set-pulse (red) and control-pulse (blue) is shown schematically in the lower panel.}
\label{Fig:6}
\end{center}
\end{figure}

As pointed out, the hysteretic behavior observed for the Q-Dot and WL emissions indicates a surprisingly slow build-up of background charge which opens the opportunity for SAW switching of the Q-Dot emission at significantly lower SAW powers compared to those required in the up-sweep. In the experiments presented so far, the SAW was applied during $1\mathrm{~\mu s}$ pulses. For the overall repetition rate of 100 kHz, the SAW is off for $90\%$ of the time. Such low duty cycles are crucially required since for the high SAW power levels necessary to switch the Q-Dot emission, unwanted heating may occur. However, if the power levels required for switching are moderate i.e. $P_{\rm ~SAW}< +15 \rm~dBm$ - as in the case of Q-Posts or in Q-Dot \emph{down}-sweeps - continuous wave (cw) SAWs can be applied without sample heating. We overcome the limitation to pulse the SAW and achieve switching for Q-Dots by applying a short, high power SAW set-pulse ($\Delta t \sim 5$ s, $P_{\rm ~SAW} > +20$ dBm). Using this set-pulse we prepare the system in a defined state by clearing the fluctuating charge environment. After this preparation step, we are able to perform switching from single line emission (no SAW) to the characteristic pattern of $1X^+$, $1X^0$ and $2X^0$ for SAW power levels as low as  to $P_{\rm ~SAW} = +10 \rm ~dBm$. The repetition time between two set-pulses can be extended to more than one minute due to the extremely slow build-up of the charge background. A typical example for Q-Dot switching with a cw-SAW is shown in \ref{Fig:6}. Here the temporal evolution of the Q-Dot emission is plotted in grayscale representation and in the lower part, the applied SAW power levels are shown schematically. A short ($5$ s) SAW prepulse with $P_{\rm ~SAW} = +20$ dBm (shown in red) is turned on at $t=0$ s and we observe the characteristic switching of the emission spectrum. Over the duration of the SAW set-pulse all emission lines show a pronounced \emph{red}-shift due to the aforementioned heating in the sample by the high power cw-SAW. After the set-pulse the SAW remains turned off for $\Delta t=20$ s and the emission spectrum switches back to a single emission line labeled $1X^{0*}$ at 1306.8 meV. Then, at $t=25$ s, the cw-SAW is switched on again with a reduced power level (control-pulse) of $P_{\rm ~SAW} = +11$ dBm (blue level in \ref{Fig:5}). Such low levels are not sufficient to achieve switching during a SAW up-sweep, however, due to the initial high power set-pulse, switching is achieved for the duration of this control-pulse. We observe the characteristic pattern for a Q-Dot consisting of the $1X^0,~2X^0$ and $1X^{+}$ emission lines at $E_{1X^0}= 1307.4$ meV, $E_{2X^0}= 1305.0$ meV and $E_{1X^+}= 1308.4$ meV, respectively. Moreover, no spectral shift  emission lines due to heating effects or blinking over the duration of the control-pulse is observed. The characteristic three line emission pattern is maintained with the cw-SAW on over 25 s until, at $t = 50$ s, the RF generator is switched off and the original emission line $1X^{0*}$ is restored. This example clearly demonstrates that reversible switching at relatively low SAW excitation can be achieved also without spurious heating effects for Q-Dots.\\ 

Our experimental findings clearly demonstrate the ability to efficiently and deliberately manipulate the charge carrier injection into self-assembled Q-Dots and Q-Posts using SAWs. This massively parallel approach allow to adjust the occupancy state of these nanostructures by the amplitude of the SAW. The large piezoelectric fields induced by the SAW control the local densities of electrons and holes in time. For high SAW amplitudes this leads inparticular to a dissociation of photogenerated electron-hole pairs which are sequentially injected into the zero-dimensional nanostructures. We exploit this effect to acousto-electrically control the probabilities for the generation of neutral and charged excitons in the two systems and increase the total emission intensity by almost one order of magnitude in the case of Q-Dots. Furthermore, we observe a pronounced narrowing of the Q-Dot emission linewidth after high SAW power levels have been applied even if carriers are excited far off-resonant. Our scheme can be readily extended to SAW-driven carrier injection into remotely located Q-Dots and Q-Posts for an acoustically triggered single photon source \cite{Wiele:98} with these high quality emitter systems. The preferential generation of neutral exciton species in Q-Posts could be further exploited for a deterministically generated biexciton state for the generation of polarization entangled photon pairs from the biexciton-exciton cascade\cite{Akopian:06}. This goal is more challenging using devices based on Coulomb blockade \cite{Benson:00} compared to our approach allowing for \emph{inherently sequential} injection of the two carrier species. Moreover, our SAW technique can be combined with planar optical microcavities\cite{Stoltz:05} to increase the photon extraction efficiency or a reduction of the emission time window due to the Purcell effect.\\

{\bf Acknowledgement:} This work was supported within the framework of the cluster of excellence "Nanosystems Initiative Munich" (\emph{NIM}), by DFG via SFB491 and SPP1285, BMBF via nanoQUIT, NSF via Nanoscale Interdisciplinary Research Team (NIRT) grant CCF-0507295, DARPA SEEDLING and by the Alexander-von-Humboldt-Foundation.\\

\bibliography{report.bib}

\section{Figures}

\end{document}